\documentclass[11pt]{article}

\usepackage[english]{babel}

\begin{document}

\begin{center}
{\large\bf On the Soviet Contribution to the Discovery of Quark
Color  \\
(Against One False Revision)}

\vspace*{0.3cm}
{\bf V.A. Petrov} \\
{\it Institute for High Energy Physics \\
142281 Protvino, Moscow Region, Russia}
\end{center}

\begin{center}
{\textbf{Abstract}}
\begin{quotation}
\noindent
A critical discussion of recent attempts to revise the modern
physics history is presented.
\end{quotation}
\end{center}

\vspace*{0.3cm}
Time after time new versions of historical records concerning
questions of priority appear. For instance, because of some
documentary findings that may disprove an established opinion.
Unfortunately, some materials are composed and displayed mostly to
cause a sensational effect, but without sufficient reason.

The present note concerns the discovery of a new quantum number
for quarks made in 1965 independently by several authors
[1a]-[1d]. The very quantum number was later dubbed ``color''. In
the USSR it was done in the beginning of 1965 by N.N. Bogoliubov,
B.V. Struminsky and A.N. Tavkhelidze [1a]. However, their paper
was not published in aregular physical journal and remained a
preprint in Russian. In the West it became known mostly as A.N.
Tavkhelidze's talk [2] at the 1965 Trieste conference on
high-energy physics and elementary particles (see, e.g., the
references in paper [3]). Nonetheless, it was generally believed
that the Soviet contribution was  made in paper [1a].

\vspace*{0.3cm}
Recently, however, an arXiv publication [4] has appeared in which
this opinion was contested in very strong terms. The main point is
as follows:
\begin{quotation}
\noindent
There exists a preprint by B. V. Struminsky alone [5] which
preceded the preprint [1a] and –- what is the most important --
which contained a footnote:

\vspace*{0.3cm}
\noindent
{\it``Three identical quarks cannot be in an antisymmetric
\mbox{S-state}. In order to form an antisymmetric \mbox{S-state}
one has to attribute  to the quark an additional quantum
number.''}
\end{quotation}

\vspace*{0.3cm}
\noindent
Being based on this and some other mostly emotional and poorly
based observations, supplemented with investigations \`{a} la
home-bred Sherlock Holmes in the JINR Library the author of [4]
came to an unambiguous conclusion:

\begin{quotation}
\noindent
\textit{\textbf {The Soviet contribution into the discovery of
color is due to solely B.V. Struminsky.}}
\end{quotation}

\vspace*{0.3cm}
In Ref. [4] this is expressed quite carefully (original
orthography):

\begin{quotation}
\noindent
{\it{``The idea that an internal quantum number for quarks,
eventually called color, can help explain the magnetic moments of
barions within the framework of standard quantum theory, was
explicitly stated in a 7th January, 1965 JINR publication by Boris
Struminsky''.}}

\vspace*{0.3cm}
\noindent
{\it {``...the symmetry argument leading to the additional quantum
number was fully and explicitly present in the booklet [1]''}} (Ref. [5])
in our references).
\end{quotation}

\vspace*{0.3cm}
\noindent
As to Ref. [1a] it  only {\it{``made an important contribution to
clarifying the dynamic aspect of the quark magnetic moments''}}. No more.

\vspace*{0.3cm}
With all this and many vague allusions and reservations the author of
Ref. [4] thoroughly brings the reader to another -- implicit but
factual -- conclusion:

\begin{quotation}
\noindent
\textit{\textbf {N.N. Bogoliubov commited an immoral deed: he
``hanged on'' Struminsky's discovery.}}
\end{quotation}

\vspace*{0.3cm} This year scientists of many countries celebrate
100-year anniversary of N. N. Bogoliubov, an outstanding and
renowned Russian/Soviet mathematician and physicist,  who
contributed so much into the most important fields of the XX-th
century mathematics and physics: nonlinear mechanics, microscopic
theory of superfluidity and superconductivity, kinetic equations
(BBGKY ierarchy), renormalization theory
(Bogoliubov-Parasiuk-Hepp-Zim\-mermann R-operation and first
consistent formulation of the renormalization group), axiomatic
field theory (Bogoliubov's system of axioms, first rigorous proof
of the dispersion relations) etc etc.  In usual competition among
physicists of different schools the sides could be sometimes quite
hostile but never a question about ethical behavior of Bogoliubov
has been raised. Nonetheless, now we are in front of rather grave
and blasphemous invectives made by the author of Ref. [4].

\vspace*{0.3cm}
This is our moral debt to give a due reply. To this end we would like
to present the following considerations.

\begin{enumerate}
\item
Bogoliubov published many papers with co-authors. In cases
he considered their own separate contributions deserving special
mentioning he did it with a perfect willingness. For instance, in
a series of papers devoted to the R-operation, he worked on
together with O.S~Parasiuk, he made a reference to the paper
written by Parasiuk alone in which the latter proved some
important theorem (see, e.g., the book ``Introduction to the
Theory of Quantized Field'' by Bogoliubov and Shirkov).

\item
According to memoirs of Bogoliubov's disciples and co-workers his
attitude towards them leaves no room for doubts in his decency.

\begin{quotation}
\noindent
{\it {``This was a very particular feature of   Nikolai Nikolaevich:
when he ``puzzled'' any of his colleagues, he always solved the
problem by himself, and later, if the results coincided, he always
said: ``But you did it better'''' [6]}}.
\end{quotation}

\item
In paper [5] B.V. Struminsky thanks his superviser Bogoliubov for
posing a problem and attention:
\begin{quotation}
\noindent
{\it {``The author expresses his sincere  gratitude to
Academician N.N.~Bogoliubov for the suggested problem and
attention''}}.
\end{quotation}

\item
In subsequent papers by Struminsky the references concerning
the invention of color were made always to paper [1a] only. One
can find in paper [1a] a reference to Struminsky's earlier paper
[5] but not in relation with the new quark quantum number  but in
relation to the quark magnetic moments only. As to the color,
Struminsky (in a joint paper with A.N.~Tavkhelidze) wrote
afterwards:
\begin{quotation}
\noindent {\it {``Another {\rm (in compare with parastatistics --
our note)} method to overcome the difficulty mentioned above was
suggested in the works of  N.N.~Bogoliubov et al. [6] {\rm ([1a]
in our references)} and those of Nambu and Han [13] {\rm ([1b] in
our references)}. The main idea of these works is the introduction
of three quark  triplets and construction of baryons from three
different quarks'' [7]}}.
\end{quotation}

Why on the earth the genuine author of a discovery decided to ascribe
it to somebody else? Why did he never express even a little bit of
doubt  in the accepted version of events?

\item
N.N. Bogoliubov, when having spoken about Soviet contribution to
the discovery of the quark color, always meant and referred to
paper [1a] and never to Ref. [5].
\end{enumerate}

A clear way out of this false problem is to acknowledge that
Bogoliubov has informed his PhD student Struminsky in general terms on
resolution of the quark statistics problem. We would like to stress
that all this in no way diminishes a scientific authority of B.V.
Struminsky himself.

\vspace*{0.3cm}
 R\'esum\'e: an attempt of a revisionistic sensation produced in Ref.
[4] has no grounds.

\vspace*{0.5cm}
\noindent
{\bf References}

\vspace*{0.3cm}
\parindent=0pt
[1a] N.N. Bogoliubov, B.V. Struminsky and A.N. Tavkhelidze,
JINR Preprint D-1968 (received 23 January 1965, in Russian); \\
N.N. Bogoliubov, V.A. Matveev, Nguen Van Hieu, D. Stoyanov,
B.V.~Struminsky, A.N.~Tavkhelidze  and  V.P.~Shelest, JINR
Preprint P-2141, Dubna, 1965 (in Russian).

\vspace*{0.3cm}
[1b] M.Y. Han and Y. Nambu, Phys. Rev. \textbf{139}
(1965) B1006 (received 12 April 1965).

\vspace*{0.3cm}
[1c] Y. Miyamoto, Progr. Theor. Phys. Suppl. Extra
number (1965) 187.

\vspace*{0.3cm}
[1d] T. Tati, Prog. Theor. Phys. \textbf{35} (1966) 126 (received
11 September 1965).

\vspace*{0.3cm}
[2] A. Tavkhelidze, \emph{Proc. of the Seminar on
High Energy Physics and Elementary Particles} (IAEA, Vienna,
Austria, 1965), p.~753.

\vspace*{0.3cm}
[3] Y. Nambu and M.-Y. Han, Phys. Rev. D \textbf{10} (1965) 674.

\vspace*{0.3cm}
[4] V.F. Tkachov, arXiv:0904.0343 (physics.hist-ph).

\vspace*{0.3cm}
[5] B.V. Struminsky, JINR preprint P-1939 (submitted on
7 January 1965, in Russian).

\vspace*{0.3cm}
[6] B.V. Medvedev, Usp. Mat. Nauk, \textbf{49} (1994) 83 (in
Russian) (Russ. Math. Surv. \textbf{49} (1994) 89).

\vspace*{0.3cm}
[7] B.V. Struminsky and A.N. Tavkhelidze, \emph{High Energy
Physics and Theory of Elementary Particles} (Kiev, 1967), p.~625
(in Russian).

\end{document}